\documentstyle[11pt,paspconf]{article}

\begin{document}

\title{Coronal activity among open cluster stars}

\author{Sofia Randich}%\altaffilmark{1}}
\affil{Osservatorio Astrofisico di Arcetri, Largo E. Fermi, 5,
    I-50125 Firenze, Italy}

\begin{abstract}
Focusing on $ROSAT$ results for clusters in the $\sim 20-600$ Myr range, 
I first summarize our current understanding of the X--ray activity -- rotation
-- age relationship. Then, the problem of the Hyades K and M dwarfs binaries is
addressed: {\it 1.} most K and M--type 
binaries in wide systems are X--ray brighter 
than single stars; {\it 2.} binaries seem to fit into the same 
activity -- rotation
relationship as single stars. Points {\it 1.} and {\it 2.} suggest that
the distributions of rotations of single and binary stars should also
show a dicothomy, but the few available rotational data do not
support the existence of such a dicothomy. Rotational periods for a larger
sample of binary and single stars should be acquired before any conclusion
is drawn. Finally, I discuss the topic whether the activity--age dependence
is unique, as commonly thought. 
Whereas the comparison of Praesepe to the Hyades might imply
that this is not the case, the X--ray activity of a sample of Hyades--aged
field stars instead supports the common thinking.

\end{abstract}

% Keywords should be included, but they are not printed in the hardcopy.


% That's it for the front matter.  On to the main body of the paper.
% We'll only put in tutorial remarks at the beginning of each section
% so you can see entire sections together.

\section{Introduction}

As an introductory remark, it is useful to recall that
X--ray emission from solar--type and lower mass stars is thought to originate
from a hot corona heated and confined by magnetic fields that are generated
through a dynamo process. It is therefore expected on theoretical grounds
that the level of X--ray emission, or coronal activity, should 
depend on at least the properties
of the convective zone, 
on stellar rotation and, through the rotation--age
dependence, on stellar age. 
X--ray surveys of stellar clusters offer a powerful tool to
empirically prove and quantitatively
constrain the dependence of coronal activity on these parameters
and, possibly, on additional ones, thus
providing feedback to the theory.
$ROSAT$ PSPC and HRI observations have provided X--ray images for about 30 
open clusters
in the age range between $\sim 20 - 600$ Myr (see Table 1 in Jeffries 1999,
for the most updated list). Our understanding of coronal properties
of solar--type and low mass stars in clusters is now considerably deeper
than a decade ago, but, at the same time, new puzzles have been raised
by $ROSAT$ results.
\begin{figure}
\vspace*{7.5cm}
\includegraphics{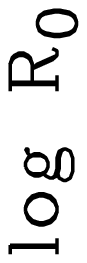}
\caption{$\log$~L$_{\rm X}$/L$_{\rm bol}$ vs. the logarithm of the Rossby
number for cluster and field stars. Open symbols are as follows; circles: 
Pleiades; squares: IC~2602 and IC~2391;
stars: Alpha Persei; triangles: Hyades single stars; crossed triangles:
Hyades binaries; diamonds: IC~4665. Filled symbols: field stars. 
The line represents the regression fit
of the points with $\log R_0 > -0.8$.}
\end{figure}

The main results
and questions emerged from $ROSAT$ observations of clusters have
been discussed in 
several reviews in the last few years.
The age -- rotation -- activity paradigm (or ARAP) has been discussed
at length by Caillault (1996), Randich (1997), and Jeffries (1999).
Other issues, such as time variability (Caillault 1996; Stern 1999;
Jeffries 1999), insights from spectra (Caillault 1996), supersaturation
(Randich 1998), 
and observational limits and analysis techniques (Micela 1996)
were also addressed. I refer to those papers for a detailed discussion
of the above topics. In the present paper I will first present a summary
of the general picture of the ARAP that we gathered from $ROSAT$ data;
second, I will address an issue that was only marginally discussed
in previous reviews, namely binaries and their influence on cluster
X--ray luminosity distribution functions (XLDF).
Finally, I will focus on the exceptions to the ARAP
and on the controversial question whether 
the X--ray properties of a cluster at a given age can be considered
as representative of all clusters at that age. Within this
context, I will compare cluster stars with field stars.

The following sources of X--ray data were used; 
{\it Pleiades}: 
Stauffer et al. (1994), Micela et al. (1996), Micela et al. (1999a);
{\it IC~2602}: Randich et al. (1995); {\it IC~2391}: Patten \& Simon (1996);
{\it Alpha Persei}: Prosser et al. (1996); {\it Hyades}: Stern et al. (1995),
Pye et al. 1994; {\it IC~4665}: Giampapa et al. (1998); {\it NGC~2547}:
Jeffries \& Tolley (1998); {\it NGC~2516}: Jeffries
et al. (1997);  {\it Blanco~1}: 
Micela et al. (1999b); {\it NGC~6475}: Prosser et
al. (1995), James \& Jeffries (1997); {\it Coma Berenices}: 
Randich et al. (1996b); 
{\it Praesepe}: Randich \& Schmitt (1995). 

\section{A consistent picture: the ARAP}

\begin{figure}
\vspace*{7.5cm}
\includegraphics{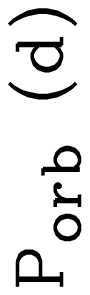}
\caption{$\log$~L$_{\rm X}$ vs. orbital period for Hyades binaries 
with B$-$V$\geq 0.8$. Crossed circles denote stars with
available measurements of rotational periods. The horizontal line represents
the median L$_{\rm X}$ of single K dwarfs.}
\end{figure}
\begin{figure}
\vspace*{7.5cm}
\includegraphics{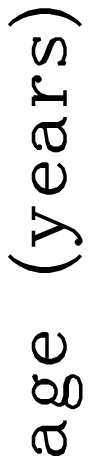}
\caption{X--ray luminosity as a function of age for solar--type stars
in clusters (circles) and in the field (diamonds). For the clusters
the median L$_{\rm X}$ is plotted. The open triangle represents
the median of a sample of Hyades--aged field stars.
The vertical lines connect the median
with the 25th and 75th percentiles. Circled symbols denote X--ray selected
cluster samples. 
The three lines represent power laws with indices $\alpha$ equal to $-0.5$ 
(dotted), $-1$ (dashed), and $-2$ (solid).}
\end{figure}
The main results evidenced by $ROSAT$ observations of open clusters
can be summarized as follows:
\begin{itemize}
\item If we exclude ``outliers" or exceptions which I will discuss in Sect.
4, the average level of X--ray activity decays with age. 
Whereas this was
already well established from {\it Einstein} observations of
the Hyades and the Pleiades (e.g., Micela et al. 1990), the larger number
of clusters observed by $ROSAT$ and the finer age sampling have 
allowed deriving
a more detailed activity vs. age relationship. The decay timescales
appear to be different for different masses (the lower the mass the longer
the timescale) and the L$_{\rm X}$ vs. age functional dependence is not
simply described by the Skumanich power law (Skumanich 1972);
\item In all clusters
the maximum X--ray luminosity (L$_{\rm X}$) decreases towards 
later spectral--types;
at a given spectral--type, a significant scatter in
L$_{\rm X}$ is observed; as a consequence, whereas the
median L$_{\rm X}$ decreases with age, the XLDFs
for clusters of different ages are not ``parallel" one to another and
some overlap is present. This means that X--ray activity cannot be
unambiguously used as an age diagnostic;
\item The X--ray activity level does depend on rotation only up to a rotation
threshold above which X--ray emission saturates; for stars rotating faster
than this threshold the ratio of the X--ray
luminosity over bolometric luminosity, L$_{\rm X}$/L$_{\rm bol}$, is about 
constant and equal to 10$^{-3}$. Note that a definitive explanation for 
saturation has not yet been offered.
\end{itemize}
$ROSAT$ observations of clusters are complemented by 
determinations of rotational velocities and/or periods in a variety
of clusters. 
Very briefly, it is now well established 
that stars arrive on the Zero Age Main Sequence
(ZAMS) with a large spread in their rotation rates and then they
slow down with mass--dependent timescales (e.g., Barnes 1999;
Bouvier 1997 and references therein).

The use of the so--called Rossby diagram 
allows incorporating the above points into a unique picture.
Noyes et al. (1984) were the first to show
that the use of the Rossby number ($R_0$),
the ratio of the rotational period P over the convective turnover time
$\tau_c$, which somehow allows formalizing
the dependence of activity on the properties of the convection zone,
improved the rotation--chromospheric activity relationship for field stars.
Randich et al. (1996a) and  Patten \& Simon  (1996) showed this to
hold also for the X--ray activity of cluster stars. 
Taking advantage of the new available periods for several clusters, I produced
an updated version of the diagram which I show in Figure~1.
X--ray data for field stars were taken from Schmitt (1997) 
and H\"unsch et al. (1998, 1999); periods were taken from Hempelmann 
et al. (1996); I retrieved periods
for most of the clusters 
from the Open Cluster Database \footnote{Open
Cluster Database, as provided by C.F. Prosser (deceased) and J.R. Stauffer,
and which currently may be accessed at http://cfa-ftp.harvard.edu/~stauffer/,
or by anonymous ftp to cfa-ftp.harvard.edu, cd /pub/stauffer/clusters.},
complementing the ones for the Pleiades with the new measurements of
Krishnamurthi et al. (1998) and adding periods for IC~2602 from Barnes et
al. (1999).
I derived Rossby numbers using 
the semi--empirical formulation for $\tau_c$ given by Noyes et al. (1984).
I refer to the paper of Pizzolato et al. (1999) for a discussion
of how different ways of estimating $\tau_c$ may affect the 
$\log \rm L_{\rm X}/\rm L_{\rm bol}$ vs. $\log R_0$ relationship.

Various features can be noted in the diagram: first, saturation of
X--ray activity is evident: it occurs at $\log R_0 \sim -0.8$. The points
with a lower Rossby number cluster around $\log$~L$_{\rm X}$/L$_{\rm bol}=-3$
(but note the supersaturation at very low $\log R_0$ --see Randich 1998).
Since the diagram includes stars from F down to M spectral--type,
the uniformity of the threshold Rossby number below which X--ray emission
is saturated, implies that the rotation threshold depends on mass.
In other words, if $\log (R_0)_{\rm thr}=(\log$~P/$\tau_c$)$_{\rm thr}
=const\sim -0.8$, then,
P$_{\rm thr} \propto \tau_c$; since $\tau_c$ increases with 
decreasing mass (the
convective envelope becomes deeper), the lower the mass, the
longer is the threshold period (e.g., Stauffer et al. 1997a).
Second, all cluster and field stars fit into a unique
relationship. This on one hand means that field and cluster stars
behave in a similar way as far as the rotation -- convection -- activity
relation is concerned; whereas this is qualitatively expected --why should
field and cluster stars behave differently?-- it is good to empirically
confirm the expectations.
On the other hand, the fact that stars in all clusters lie on the same
curve, irrespective of age and mass, implies that the activity--age dependence
is most likely an activity--[rotation--convection]--age dependence. 
Incidentally, whereas a certain amount of scatter 
around the relation is present, as well as a few outliers,
I believe, in agreement with 
Jeffries (1999), that most of the scatter is likely due to errors and
non--uniformity in L$_{\rm X}$ measurements and to some variability in
X--ray luminosities.
Third, the linear regression curve has a slope
equal to $-2.1(\pm 0.09)$ which, at a given spectral--type
(i.e, roughly constant $\tau_c$ and stellar radius) is the same functional
L$_{\rm X}$ vs. rotational velocity dependence
found by Pallavicini et al. (1981) for field stars.

In summary, the Rossby diagram can be looked at as an evolutionary diagram.
Stars arrive on the ZAMS characterized by a range of
rotation rates; therefore they occupy different regions of the Rossby diagram,
with a significant
part of them lying on the saturated part. The maximum luminosity
at a given spectral--type is bounded by the saturation condition which explains
why it decreases towards late spectral--types;
non--saturated stars cause the spread in L$_{\rm X}$, whilst
saturated stars, in principle, do not contribute to it.
As the clusters age, the stars spin--down
and they move towards the right of the Rossby diagram. Their L$_{\rm X}$ remain
virtually unchanged until they de-saturate and, once they do not lie anymore
on the saturation plateau, they become progressively less active as they 
continue to spin-down. 
The fraction of saturated
stars in a cluster decreases until, as is the case for the Hyades
solar--type stars, all the stars are non--saturated. As a consequence,
the mean and median luminosities decay. Since, as we consider later 
spectral--types,
both the spin--down timescales and the saturation threshold
period are longer, K and M dwarfs move towards the right of the diagram
at a slower rate than solar--type stars (in other words, they remain 
saturated longer); accordingly,
the timescales for the decay of X--ray activity  of K and M dwarfs are 
also longer than for solar--type stars. 

\section{Binaries}

How do binary stars fit into the scenario outlined in the previous section?
In principle, there should be no difference between single stars and wide
binaries, which, therefore, should follow a ``normal" X--ray activity -- 
rotation -- age
evolution. On the contrary, as well known,
binaries in close, tidally locked systems, are
rapid rotators even at rather old ages and therefore are expected to
show high levels of X--ray activity and to contribute
to the high luminosity tail of a cluster XLDF.

In young clusters like the Pleiades, virtually no difference is observed
between the X--ray activity level of single and binary stars (e.g., Stauffer et
al. 1994);
this is indeed not surprising since most of the Pleiades single stars are
still rapid rotators because of their young age.

The situation is different in the older Hyades: the X--ray brightest stars 
in the cluster are well known binaries. Most surprisingly, however,
not only tidally locked BY Dra binary systems 
are found to be more active
than single stars, but a high level of X--ray emission is also observed among
several wide binaries.
The influence of binary systems on the XLDFs of the Hyades has been discussed
by Pye et al. (1994), Stern et al. (1995), and Stern and Stauffer (1996).
All these studies pointed out that the XLDFs of late--A, F, and G--type binaries
are very similar to those of single stars.
On the contrary, the XLDFs of binary and single K and M dwarfs show a dicothomy,
with the bulk of the binary population being considerably
more X--ray active than single stars (see Fig.~10 in Stern et al. 1995 and 
Fig.~2b of Pye et al. 1994). Pye et al. estimated that the 
probability that binary and single K--type stars XLDFs are drawn from
the same parent population is lower than 0.4 \%.
Since most of the K--type binaries are
in wide systems with orbital periods of the order of a year or longer, 
enforced rotation could not be the reason for the high
activity level. Pye et al. also showed that the higher luminosities
of binary K dwarfs could not simply be due to the summed luminosities
of single components. 

The questions then arise {\it a)} whether the rotation--activity
relationship for binaries is similar to that of single stars and, {\it b)}
if this is 
the case, why do binaries in long period systems maintain high rotation
and activity. Hyades binaries with known rotational periods are plotted
as crossed triangles in the Rossby diagram shown in Fig.~1; they clearly
follow the same $\log R_0$ vs. L$_{\rm X}$/L$_{\rm bol}$
relation as single Hyades stars,
with only one binary lying above the locus of the other stars
(the star is VB~50, B$-$V$=0.59$ --i.e., it is not a K/M--type binary).
The answer to question {\it a)} seems therefore to be ``yes".

Figure~2 is a revised version of Fig.~11 of Stern et al. (1995); in the
figure I plot the logarithm of X--ray luminosity
as a function of the orbital period (P$_{\rm orb}$) for Hyades binaries
with B$-$V $\geq 0.8$. Orbital periods come from
various sources in the literature and were retrieved from the Open Cluster
Database. The figure indeed confirms that most wide binaries have a higher
L$_{\rm X}$ than the median luminosity of single stars. Stars with 
P$_{\rm orb} \leq 10$ days are synchronous, 
as expected (e.g., Zahn \& Bouchet 1989) and they nicely follow a L$_{\rm X}$
vs. P$_{\rm orb}=\rm P_{\rm rot.}$ relationship
(in agreement with the trend seen in Fig.~1). The stars with longer
orbital periods do not follow such a relationship, but are scattered
throughout the diagram. Only three of them have available rotational
period, but for these three stars
a L$_{\rm X}$ vs. P$_{\rm rot}$ relationship may also hold, with the most
active one being the most rapid rotator.
In other words, both Figs.~1 and 2 suggest
that rotation is the reason for the high activity level of 
both short--period and long period binaries and that even binaries in wide
systems may maintain a rather high rotation (at least higher than single stars).

As possible explanations for this Pye et al. (1994) and Stern et al. (1995)
proposed either the higher initial angular momentum available in binary systems
or a different PMS rotational evolution; more specifically, the reasonable
hypothesis could be made that binaries
disrupt their circumstellar disks earlier than single stars, thus 
removing a source of rotational braking. If this is the case, as stressed
by Stern \& Stauffer (1996) the rotational velocity distributions of single
and binary K and M dwarfs should also show a dicothomy. Contrarily to
this expectation Stauffer et al. (1997b), based on v$\sin i$ measurements,
found that the components of SB2 
binaries in the Hyades are, on average, slow rotators.

In summary, we are left with the contradicting evidences that {\bf 1.} the
same L$_{\rm X}$ vs. period or $R_0$ relationship holds for binaries and
single stars; {\bf 2.} wide K and M dwarfs binaries may exist with
rather short rotational periods and high activity levels; {\bf 3.} 
the v$\sin i$
distributions of the sample of K and M--type binaries and single stars 
studied by Stauffer et al. (1997b) do not show any evident dichotomy.
I think two possible reasons for this inconsistency can be proposed;
first neither the sample of wide 
binaries with known orbital and rotational periods,
nor the sample of Stauffer et al. (1997b) are large enough, and more
important, complete. Second, rotational periods of $\sim 10$ days correspond,
for stars with B$-$V $\sim 0.9$ (see Fig.~2) to velocities of the order
of 4 km/s, lower than 
the v$\sin i=6$ km/s detection limit of Stauffer et al. (1997b); this suggests
that the dicothomy between single and binary K and M dwarfs may show up only
among slow rotators.
Rotational periods for a large sample of both binary and single stars
are clearly required to further investigate this issue.

\section{Problems with the ARAP}

As discussed in Sect.~2, most of the $ROSAT$ results for open clusters
can be explained within the ARAP scenario. Whereas the Rossby diagram
shown in Fig.~1 evidences  no major deviations from the activity--rotation
relationship, exceptions to the age--activity relationship have instead
been found. I focus here on solar--type stars, but I mention that problems
also exist for lower mass stars.

In Figure~3 I plot the median L$_{\rm X}$ vs. age for G--type stars (0.59
$\leq$ B$-$V$_0 \leq 0.82$) in various clusters. The vertical bars denote
the luminosity range between the 25th and 75th percentiles of the XLDFs.
Field stars are also included in the plot. Their age was taken from Ng \&
Bertelli (1998) or Edvardsson et al. (1993): all but one are older than the 
Hyades. The open triangle 
indicates the median luminosity of a sample of nine field stars with an 
age similar to the
Hyades; I selected these stars using lithium measurements from Pasquini
et al. (1994), under the plausible assumption that Li in this color range
and up to the Hyades age is
a reliable age indicator. 
Three lines denoting power laws with indices $\alpha=
-0.5$ (Skumanich law), $-1, -2$ are also shown in the diagram.

The figure witnesses the general trend of decreasing X--ray emission with 
increasing age, the fact that the decay cannot be simply described
by a power law, and the overlap between XLDFs of different clusters
(i.e., the most active Hyades stars can be as active as stars in the
Pleiades). Not all the clusters, however, fit into the mean trend: 
Praesepe appears
to be the most 
discrepant cluster in the diagram. It has about the same age as the Hyades
and Coma, but as the figure shows, the bulk of its solar--type stars
population is considerably X--ray fainter than the other two clusters
(Randich \& Schmitt 1995). Barrado y Navascu\'es et al. (1998) demonstrated
that such a result is not due to the contamination 
by non--members in the Praesepe sample. In addition,
according to Mermilliod (1997), the distributions of rotational
velocities in the Hyades and Praesepe are rather similar, although v$\sin i$ or
periods are not currently published and thus it is not possible to check
on a star-to-star basis whether Praesepe stars follow the same 
activity -- rotation relationship as the stars in other clusters. 
In any case, this discrepancy casts doubts on the assumption that the
X--ray properties of a cluster at a given age can be considered as 
representative
of all clusters at that age. Totten et al. (1999)
and Franciosini et al. (1999) analyzed a $ROSAT$ HRI image of
NGC 6633, a cluster of about the same age as the Hyades and Praesepe: 
both studies
found that NGC~6633 seems to be more Praesepe--like than Hyades--like,
supporting the conclusion that the age--activity relation is not unique
(but deeper X--ray observations of NGC~6633 are needed to confirm that
NGC~6633 is really less active than the Hyades). On the contrary,
as Fig.~3 shows, the median X--ray luminosity of a random sample of
field stars at $\sim$ 600~Myr exactly matches the Hyades median (and the
spread around the median is very small), supporting 
the opposite conclusion that the Hyades are indeed the standard 
at 600~Myr. A solution to this puzzle (at least as far as the
Hyades/Praesepe dichotomy is concerned) is possibly offered by the results
of Holland et al. (1999) who suggest that Praesepe could result from two
merged clusters, with the brightest X--ray sources being found almost 
exclusively in the main cluster.

Other (minor) inconsistencies are visible in Fig.~3; whereas it is understood
why all clusters up to Alpha Persei have about the same median luminosity
(there is no substantial spin--down up to that age), a tight age--activity
relationship does not appear to
hold between $\sim$100 and 250 Myr. This, again, would
imply that the age--activity relationship is not unique and that
other parameters (metallicity? e.g., Jeffries et al. 1997)
besides rotation and age influence the level of X--ray
activity. However, several sources of uncertainty should be removed
before such a conclusion can be regarded as definitive. Namely: {\it i)}
the X--ray data used to compute XLDFs and the median luminosities come
from different surveys, with different sensitivies and have been analyzed
in different ways (I just used the published X--ray luminosities);
{\it ii)} some of the cluster samples are X--ray selected, and thus
biased toward X--ray bright stars; {\it iii)} some of the cluster samples
(e.g., Blanco~1) may be contaminated by non--members; {\it iv)}
the clusters shown in the figure are not on the same age scale; whereas
ages for the Pleiades, Alpha Persei, and IC~2391 come from the most
recent determinations through the lithium boundary method, the ages
for the other clusters are the more traditional ones derived through
color--magnitude diagram fitting. Note, for example, that the age of NGC~2547
could indeed be larger (see Jeffries et al. 1999).

Finally, the X--ray activity--age relation for stars older than the Hyades
is defined by field stars only, which are scattered throughout the diagram. 
The figure may suggest that the decay between the Hyades and e.g., the Sun
is more rapid than $t^{-1/2}$, but, very obviously,
X--ray surveys deep enough to reach main sequence solar--type
stars in clusters older than the Hyades are needed.

\section{Conclusions}

$ROSAT$ observations of clusters have increased our confidence in the
ARAP, but, at the same time, have led to results that cannot apparently
be fully explained by the ARAP. Before the conclusion is drawn that exceptions
to the ARAP really exist, additional X--ray and optical observations
should be carried out. The need for X--ray surveys of clusters older 
than the Hyades or of deeper observations of clusters that have already been
observed by $ROSAT$ is
unquestionable. At the same time, X--ray spectra of cluster stars will allow us
to infer their coronal properties and follow their evolution with age, or
will possibly provide us with a key to the understanding of saturation and
supersaturation. 
I refer to Jeffries (1999) for a detailed list of the issues that the
capabilities of XMM and Chandra will allow us to address.

I would like to stress here that complementary optical data (i.e., additional
determinations of periods, rotational and radial velocities, deep imaging; etc.)
are also needed
in order to address in detail these issues and, possibly, find a
solution to the puzzles discussed in the previous sections.

% Finally, we have a little acknowledgements section.

\acknowledgments
I am grateful to Giusi Micela and Roberto Pallavicini for their 
careful reading of the
manuscript and useful comments. I thank Rob Jeffries for anticipating
his results on the age of NGC~2547. This work has made extensive use
of the {\sc simbad} database maintained by the Centre
de Donn\'ee Astronomiques de Strasbourg.

% That's the end of the main body of the paper.  Now we will have some
% back matter.

% Now comes the reference list.  Since we typed out the citations ourselves,
% the reference list is enclosed in a "references" environment.  Each
% new reference begins with a \reference command which sets up the proper
% indentation.  Typography that may be required in the reference list by
% the editorial staff must be included by the author.
%
% Observe the "standard" order for bibliographic material: author name(s),
% publication year, journal name, volume, and page number for articles.
% Some journal names are available as macros; see the WGAS markup
% instructions for a listing of which ones have been "macro-ized".
% Note the use of curly braces to delimit the font changes: it is essential
% that this be done to limit the scope of the font declaration.
%
% There is no need to engage in any other typographic manipulation.


\begin{references}
\reference Barnes, S.A. 1999, these Proceedings
\reference Barnes, S.A., et al. 1999, ApJ, 516, 263
\reference Bouvier, J. 1997, Mem. SaIt, 68, 881
\reference Barrado y Navascu\'es, D., Stauffer, J.R., and Randich, S. 1998,
ApJ, 506, 347
\reference Caillault, J.-P. 1996,
The Ninth Cambridge Workshop on Cool Stars,
Stellar Systems and the Sun, R. Pallavicini and A.K. Dupree (eds), ASP
Conference Series 109, p. 325
\reference Edvardsson, B., et al. 1993, A\&A, 275, 101
\reference Franciosini, E., Randich, S., and Pallavicini, R.
1999, these Proceedings
\reference Giampapa, M.S., Prosser, C.F., and Fleming, T.A. 1998, ApJ, 501, 624
\reference Holland, K., et al. 1999, these Proceedings
\reference Hempelmann, A., Schmitt, J.H.M.M., and Stcepien, K. 1996, A\&A 305,
284
\reference H\"unsch, M., Schmitt, J.H.M.M., and Voges, W. 1998,
A\&AS, 132, 155
\reference H\"unsch, M., et al. 1999, A\&AS, 135, 319
\reference James, D.J., and Jeffries, R.D. 1997, MNRAS, 292, 252
\reference Jeffries, R. 1999, in Solar and Stellar Activity: Similarities
and Differences, C.J. Butler and J.G. Doyle (eds), ASP Conference Series 158,
p. 75
\reference Jeffries, R.D., and Tolley, A.J. 1998, MNRAS, 300, 331
\reference Jeffries, R.D., Thurston, M.R., and Pye, J.P. 1997, MNRAS, 287, 501
\reference Jeffries, R.D., et al. 1999, these Proceedings
\reference Krishnamurthi, A., et al. 1998, ApJ, 493, 914
\reference Mermilliod, J.-C. 1997, Mem. SaIt, 68, 859
\reference Micela, G. 1996,
The Ninth Cambridge Workshop on Cool Stars,
Stellar Systems and the Sun, R. Pallavicini and A.K. Dupree (eds), ASP
Conference Series 109, p. 347
\reference Micela, G., et al. 1990, ApJ, 348, 557
\reference Micela, G., et al. 1996, ApJS, 102, 75
\reference Micela, G., et al. 1999a, A\&A, 341, 751
\reference Micela, G., et al. 1999b, A\&A, 344, 83
\reference Ng., Y.K., and Bertelli, G. 1998, A\&A, 329, 943
\reference Noyes, R.W., et al. 1984, ApJ, 279, 763
\reference Patten, B.M., and Simon, T. 1996, ApJS, 106, 489
\reference Pallavicini, R., et al. 1981, ApJ, 248, 279
\reference Pasquini, L., Liu, Q., and Pallavicini, R. 1994, A\&A 287, 191
\reference Pizzolato, N., et al. 1999, these Proceedings
\reference Prosser, C.F., et al. 1995, AJ, 110, 1229
\reference Prosser, C.F., et al. 1996, AJ, 112, 1570
\reference Pye, J.P., et al. 1994, MNRAS, 266, 798
\reference Randich, S. 1997, Mem. SaIt, 68, 971
\reference Randich, S. 1998, The Tenth Cambridge Workshop on Cool Stars,
Stellar Systems and the Sun, R.A. Donahue and J.A. Bookbinder (eds), ASP
Conference Series 154, p. 501
\reference Randich, S., and Schmitt, J.H.M.M. 1995, A\&A, 298, 115
\reference Randich, S., et al. 1995, A\&A, 300, 134
\reference Randich, S., et al. 1996a, A\&A, 305, 785
\reference Randich, S., Schmitt, J.H.M.M., and Prosser, C.F. 1996b
A\&A, 313, 815
\reference Schmitt, J.H.M.M. 1997, A\&A, 318, 215
\reference Skumanich, A. 1972, ApJ, 171, 565
\reference Stauffer, J.R., et al. 1994, ApJS, 91, 625
\reference Stauffer, J.R., et al. 1997a, ApJ, 479, 776
\reference Stauffer, J.R., et al. 1997b, ApJ, 475, 604
\reference Stern, R.A. 1999, in Solar and Stellar Activity: Similarities
and Differences, C.J. Butler and J.G. Doyle (eds), ASP Conference Series 158,
p. 47
\reference Stern, R.A., and Stauffer, J.R. 1996, The Ninth
Cambridge Workshop on Cool Stars, Stellar
Systems and the Sun, R. Pallavicini and A.K Dupree (eds), ASP Conference
Series 109, p. 387
\reference Stern, R.A., Schmitt, J.H.M.M., and Kahabka, P.T. 1995, ApJ, 
448, 683
\reference Totten, E.J., et al. 1999, these Proceedings
\reference Zahn, J.-P, and Bouchet, L. 1989, A\&A, 223, 112

\end{references}
\end{document}